\documentclass[prc,twocolumn,showpacs,loatfix,nofootinbib,preprintnumbers]{revtex4}
\usepackage{txfonts}
\usepackage{amssymb}
\usepackage[section] {placeins}
\usepackage{dcolumn}
\usepackage{graphicx}% Include figure files
\usepackage{dcolumn}% Align table columns on decimal point
\usepackage{bm}% bold math
\usepackage{latexsym}
\usepackage{feynmf}     %Package for feynman diagrams.
\usepackage{longtable}
\usepackage{color}

\usepackage{ulem}
\usepackage[toc,page,title,titletoc,header]{appendix}
\usepackage[colorlinks,
            linkcolor=blue,
            anchorcolor=blue,
            citecolor=blue
            ]{hyperref}

\begin{document}
\def\bsb{{\hbox{\boldmath $\beta$}}}
\def\undersim#1{\setbox9\hbox{${#1}$}{#1}\kern-\wd9\lower
    2.5pt \hbox{\lower\dp9\hbox to \wd9{\hss $_\sim$\hss}}}
\def\brho{{\hbox{\boldmath $\rho$}}}
\def\bbox#1{\hbox{\boldmath${#1}$}}
\def\gtsim{$\raisebox{0.6ex}{$>$}\!\!\!\!\!\raisebox{-0.6ex}{$\sim$}\,\,$}
\def\ltsim{$\raisebox{0.6ex}{$<$}\!\!\!\!\!\raisebox{-0.6ex}{$\sim$}\,\,$}
\def\pt{\bbox{p}_t}
\def\rr{\hbox{\boldmath{$ r $}}}
\def\pp{\hbox{\boldmath{$ p $}}}
\def\pt{\pp_{{}_T}}
\def\yb{{\bar y }}
\def\mt{m_{{}_T}}
\def\zb{{\bar z }}
\def\tb{{\bar t }}
\def\rhoe{ {\rho_{\rm eff}} }
\def\qq{\hbox{\boldmath{$ q $}}}

\title{Long-range azimuthal correlations for partially coherent pion emission in proton-proton collisions}

\author{Peng Ru$^{1,\,2,\,3,}$\footnote{Electronic address:
p.ru@m.scnu.edu.cn}}
\author{Wei-Ning Zhang$^{3,\,4,}$\footnote{Electronic address:
wnzhang@dlut.edu.cn}}

\affiliation{$^1$Guangdong Provincial Key Laboratory of Nuclear Science, Institute of Quantum Matter,
South China Normal University, Guangzhou 510006, China\\
$^2$Key Laboratory of Quark $\&$ Lepton Physics (MOE) and Institute of Particle Physics,
Central China Normal University, Wuhan 430079, China\\
$^3$School of Physics, Dalian University of Technology, Dalian, Liaoning 116024, China\\
$^4$Department of Physics, Harbin Institute of Technology, Harbin, Heilongjiang 150006,
China}

%\received{\today}

\begin{abstract}
In this paper, we investigate the influence of coherent pion emission on long-range azimuthal
correlations in relativistic proton-proton collisions. We study the pion momentum distribution for
a coherent source with both transverse and Bjorken longitudinal expansions, and calculate the two-particle
correlation function $C(\Delta\eta,\Delta\phi)$. A ``ridge" structure is observed in the correlation
function $C(\Delta\eta,\Delta\phi)$ of the coherent pion emission, whether the coherent source is
expanding or static. The onset of this long-range azimuthal correlation can be traced back to the
asymmetric initial transverse profile of the coherent source, owing to the interference in coherent emission.
We further construct a partially coherent pion-emitting source by incorporating a chaotic emission component.
From the experimental data of the two-pion Hanbury-Brown-Twiss correlations in pp collisions at $\sqrt{s}\!=\!7$~TeV,
we extract a coherent fraction of pion emission, which increases with increasing charged-particle multiplicity,
as an input of the partially coherent source model. The ridge structure is observed
in the correlation function $C(\Delta\eta,\Delta\phi)$ for the partially coherent source.
The correlation becomes stronger at higher multiplicities because of the larger degrees of coherence.
The results in this work are meaningful for fully understanding the collectivity in the small system of pp collisions.
\end{abstract}

\maketitle

\section{Introduction}
\label{introduction}
Collectivity is an important issue in high-energy nuclear physics, about which the long-range, near-side
enhancement in the two-particle angular correlation function is one of the iconic phenomena.
Such a ``ridge" structure was first observed in relativistic nucleus-nucleus~(AA)
collisions~\cite{Adams:2005ph,Alver:2009id,Chatrchyan:2011eka,ATLAS:2012at}, and is widely regarded as a
product of the collective dynamics of the created hot/dense quantum chromodynamic~(QCD) system~\cite{Romatschke:2007mq,Song:2007fn,Schenke:2010rr,Bozek:2011ua,Gardim:2012yp,Gale:2012rq,Song:2013qma,
Pang:2013pma,He:2015hfa,Song:2017wtw}. However, a similar phenomenon later observed in high-multiplicity
proton-nucleus~(pA) and proton-proton~(pp) collisions has aroused new discussions on the origin of the
collectivity in such small systems~\cite{Khachatryan:2010gv,ABELEV:2013wsa, Khachatryan:2015lva, Aad:2015gqa,Aaij:2015qcq,Aaboud:2016yar,Khachatryan:2016txc,PHENIX:2018lia,Adam:2019woz}.
Theoretical progress based on various mechanisms involving initial- and final-state physics has been
made~\cite{Dumitru:2010iy, Werner:2013ipa,Bozek:2013ska,Ma:2014pva,Bzdak:2014dia,Schenke:2016lrs,Sanchis-Lozano:2016qda,Weller:2017tsr,
Greif:2017bnr,Bierlich:2017vhg,Zhao:2017rgg,Mace:2018vwq,Blok:2018xes,Zhang:2019dth,Heinz:2019dbd,Nie:2019swk,Schenke:2019pmk}.
However, much work remains to be done to fully understand the collectivity in the small systems.

Hanbury-Brown-Twiss~(HBT) interferometry of identical bosons~(e.g., pions) can provide
insight into the geometry and coherence of particle emission in high-energy nuclear and hadronic
collisions~\cite{Gyulassy:1979yi,Wiedemann:1999qn,Weiner:1999th,Lisa:2005dd}.
As is well known, HBT correlations arise in a chaotic~(incoherent) particle emission and will be suppressed
with the presence of coherence in particle emission~\cite{Akkelin:2001nd,Wong:2007hx,Liu:2013wlv,Liu:2014ina,Gangadharan:2015ina,Bary:2018sue}.
This is valuable for tracing the origin of collectivity. In a previous work~\cite{Ru:2017nkc}, it was
shown that the particle elliptic anisotropy, $v_2$, can be generated separately in chaotic and coherent emissions
in different mechanisms. In chaotic emission, as in a hydrodynamical model, $v_2$ can
be developed in the collective expansion of the particle source, whereas in coherent emission it is
established through the interference effect~\cite{Ru:2017nkc}, and is primarily a quantum-mechanical
response to the source configuration~\cite{Anderson:1995gf,Davis:1995pg}.

It is noteworthy that a significant suppression of two-pion HBT correlation strength was recently
observed in small systems~\cite{Aamodt:2011kd,Khachatryan:2011hi,Aad:2015sja,Aaij:2017oqu,Sirunyan:2017ies},
indicating that there is probably a substantial coherent fraction in the pion emission in such an
environment~\cite{Glauber:2006gd}. Compared to AA collisions, the effect of coherence may survive
more easily in pA and pp collisions due to the less important hot-medium surroundings. Therefore,
it will be interesting to see whether the long-range azimuthal correlations can exist in coherent
particle emission, and how this effect will influence the collectivity in small systems.

In this paper, we investigate the influence of coherent pion emission on the long-range azimuthal
correlations in relativistic pp collisions. We study the pion momentum distribution for a coherent source
with both transverse and longitudinal expansions, and calculate the two-particle correlation
function $C(\Delta\eta,\Delta\phi)$. We observe a ridge structure in the correlation function $C(\Delta\eta,\Delta\phi)$.
We further construct a partially coherent pion-emitting source by incorporating the coherent source
and a chaotic emission source described with the blast-wave model. From the ATLAS data of the two-pion
HBT correlations in pp collisions at $\sqrt{s}\!=\!7$~TeV, we extract a coherent fraction of pion emission as an
input of the partially coherent source model. We find that the long-range azimuthal correlation for the
partially coherent source becomes stronger with increasing multiplicity.

The rest of this paper is organized as follows. In Sec.~\ref{coherent}, we study the long-range azimuthal correlations
for coherent pion emission. We review the general expressions of the pion momentum distribution of coherent
emission and formulate the coherent pion-emitting source with both transverse and longitudinal expansions.
We then calculate the two-particle correlation function. In Sec.~\ref{part_coh}, we focus on the partially
coherent emission of pion in pp collisions at $\sqrt{s}\!=\!7$~TeV. We extract the coherent fraction in pion
emission from the experimental data of HBT correlations and discuss the results of the partially coherent source model.
Finally we give a summary and discussion in Sec.~\ref{conclusion}.

\section{Long-range azimuthal correlations for coherent pion emission}
\label{coherent}
\subsection{Pion momentum distribution for coherent emission}
\label{CS}
The radiation field of a classical source~(current), as is well known, is a perfectly coherent
multi-particle system~\cite{Gyulassy:1979yi,Glauber:1951zz,Glauber:1962tt,Glauber:1963fi,Glauber:1963tx}.
In this scenario, the final state of the pion field produced by a classical source $\rho(X)
\!\equiv\!\rho(t, \rr)$ is a coherent state written as~\cite{Gyulassy:1979yi}
\begin{equation}
|\phi_{\pi}\rangle =e^{-\bar{n}/2}\exp\left( i\! \int \! d^3\!p \,{\cal A}(\pp)\,
a^\dag(\pp) \right)|0\rangle,
\end{equation}
where $a^\dag(\pp)$ is the pion creation operator for momentum $\pp$, and ${\cal A}(\pp)$ can be
interpreted as the amplitude for the classical source to emit a pion with momentum $\pp$~\cite{CYWong_book}
expressed with the on-shell ($E^2_p=\pp^2+m_{\pi}^2$) Fourier transform of $\rho(X)$ as
\begin{eqnarray}
\label{FTclas}
&&{\cal A}(\pp)=A(\pp)\!\int\!d^4X\,e^{i(E_p t-\pp\cdot\rr)}\rho(t,\rr),\\ \cr
%\equiv A(\pp)\,\tilde{\rho}(\pp),\\ \cr
\label{cohpoint}
&&\,\,\,\,\,\,\,\,\,\textrm{with}\,\,\,\,\,A(\pp)=\left[2E_p(2\pi)^3\right]^{-1/2}
\end{eqnarray}
corresponding to the pion emission amplitude of a point-like source.
In addition, the pion number for the coherent state $|\phi_{\pi}\rangle$ obeys a Poisson distribution
with a mean $\bar{n}\!=\!\int \!d^3\!p\left|{\cal A}(\pp)\right|^2$~\cite{Gyulassy:1979yi}.

The most useful property for the coherent state $|\phi_{\pi}\rangle$ is that it is an eigenstate
of the annihilation operator, namely
\begin{equation}
\label{eigen}
a(\pp)\,|\phi_{\pi}\rangle=i{\cal A}(\pp)\,|\phi_{\pi}\rangle.
\end{equation}
With this property the single-pion momentum distribution can be written as
\begin{eqnarray}
&&P_{\!_C}(\pp)
\equiv\frac{d^3\bar{n}}{d^3p}
={\rm Tr}\left[D_{\pi}\,a^\dag\!(\pp)\,a(\pp)\right]
=\left|\,{\cal A}(\pp)\,\right|^2,
%=\left|\,A(\pp)\,\tilde\rho(\pp)\,\right|^2,~~~~
\label{P_c}
\end{eqnarray}
with $D_{\pi}\!\equiv\!|\phi_{\pi}\rangle\langle\phi_{\pi}|$ being the density matrix of the coherent state.
The amplitude ${\cal A}(\pp)$ expressed in Eq.~(\ref{FTclas}) can be viewed as a coherent
superposition of the pion-emission amplitudes at different source points~\cite{Ru:2017nkc,CYWong_book}.
From this perspective, the observable $P_{\!_C}(\pp)$, as the square of the absolute value of
${\cal A}(\pp)$, is linked up with the space-time geometry of the pion-emitting source $\rho(X)$
through an interference effect.

Utilizing Eq.~(\ref{eigen}), the multi-pion momentum distribution can be written as the product
of the single-pion distributions,
\begin{eqnarray}
&&P_{\!_C}(\pp_1,\dots,\pp_m)\!=\!{\rm Tr}\big[D_{\pi}\,a^\dag\!(\pp_1)\cdots
a^\dag\!(\pp_m)\,a(\pp_m)\cdots a(\pp_1)\big]~~~~~\nonumber\\
&&\hspace*{21mm}=\left|\,{\cal A}(\pp_1)\,\right|^2 \cdots\left|\,{\cal A}(\pp_m)\,
\right|^2.
\end{eqnarray}
This factorization property underlies the absence of the HBT effect in
a coherent state~\cite{Glauber:1963fi}.

As a comparison, the momentum distributions for chaotic particle emissions such
as thermal emissions are quite different~\cite{Cooper:1974mv}. For example, the single-particle momentum distribution
of thermal emission mainly depends on the source temperature, and in part can be connected to
the source geometry through a collective dynamical evolution~(flow effect) rather than interference,
due to the independent particle emissions at different source points~\cite{Ru:2017nkc,CYWong_book}.
Moreover, the multi-particle momentum distribution for chaotic emission cannot be factorized into the product of the
single-particle distributions, which gives rise to the HBT correlations.

In relativistic heavy-ion collisions and proton-proton collisions, the pion-emitting source is
possibly partially coherent~\cite{Glauber:2006gd}. Since there is no interference effect between the chaotic
and coherent emissions in the single-particle momentum distribution~\cite{CYWong_book}, the total distribution for
a partially coherent pion source can be written as
\begin{eqnarray}
P\,(\pp)=f_c\,P_{\!_C}(\pp)+(1\!-\!f_c)\,P_{\chi}(\pp),
\label{parcoh}
\end{eqnarray}
where $P_{\!_C}(\pp)$ and $P_{\chi}(\pp)$ are the normalized distributions~($\int d^3pP_{\!_C/\chi}(\pp)\!=\!\int d^3pP\,(\pp)$)
for the coherent and chaotic emissions, respectively, and $f_c$ represents the coherent fraction of pion emission.
Detailed discussions on the two- and multi-pion momentum distributions as well as the related HBT
correlations for partially coherent source can be found in Refs.~\cite{Wong:2007hx,Gangadharan:2015ina,Bary:2018sue}.
In general, the strength of HBT correlation will decrease with an increasing coherent fraction of pion emission.

\subsection{Coherent pion-emitting source with transverse and longitudinal expansions}
\label{expansion}
In the context of high-energy nuclear and hadronic collisions, it is meaningful to deliberate
the relativistic expansion of the coherent pion-emitting source.
In a previous work~\cite{Ru:2017nkc}, the effect of the transverse expansion of the coherent
source was studied. To further address observables such as long-range azimuthal correlations,
it is essential to properly consider the longitudinal structure of the coherent source.
In this work, we study a coherent source undergoing a Bjorken longitudinal expansion~\cite{Bjorken:1982qr}.

It is convenient to write the initial space-time distribution of such an evolving source in
the ${\cal X}\!\equiv\!(\tau, x,y,\eta_{s})$ representation rather than in the $X\!\equiv\!(t,\rr)\!\equiv\!(t, x, y, z)$ one,
where $\tau\!=\!\sqrt{t^2-z^2}$ is the longitudinal proper time and $\eta_s\!=\!(1/2)\ln[(t+z)/(t-z)]$ is the
space-time rapidity~\cite{Bjorken:1982qr}. We consider an initial distribution of the coherent source written as
\begin{eqnarray}
\rho_{\rm init}({\cal X}_0)=\rho_{\rm init}(\tau_0, x_0,y_0,\eta_{s0})
\,\,\,\,\,\,\,\,\,\,\,\,\,\,\,\,\,\,\,\,\,\,\,\,\,\,\,\,\,\,\,\,
\,\,\,\,\,\,\,\,\,\,\,\,\,\,\,\,\,\,\,\,\,\,\,\,\,\,\,\,\,\, \cr\cr
=\frac{(R_xR_y)^{-1}}{4\pi\Delta\eta_{s0}}\exp{\left(-\frac{x_0^2}{2R_x^2}\!
-\!\frac{y_0^2}{2R_y^2}\right)}~\textrm{Rect}\left(\frac{\eta_{s0}}{2\Delta\eta_{s0}}\right) ~\delta(\tau_0-\tau_i),
\label{GausS}
\end{eqnarray}
where $R_x$ and $R_y$ represent the sizes of the Gaussian transverse profile at an initial time
$\tau_i$. The coherent source is assumed to be initialized in a finite space-time rapidity range
$[-\Delta\eta_{s0},\Delta\eta_{s0}]$, formulated as the rectangle function.

To take into account both the transverse and longitudinal expansions, we further assume that
each of the source elements has a velocity $\textbf{\textit{v}}\!=\!(v_x,v_y,v_z)$ in the source
center-of-mass frame~(CMF) written as~\cite{Ru:2017nkc,Zhang:2006sw,Yang:2013mxa}
\begin{eqnarray}
&&v_x({\cal X}_0)=\left(\cosh\eta_{s0}\right)^{-1}\textrm{Sign}(x_0)\cdot a_x\left(\frac{|x_0|}{R_{x,\,\textrm{\small{max}}}}\right)^{b_x},\cr\cr
&&v_y({\cal X}_0)=\left(\cosh\eta_{s0}\right)^{-1}\textrm{Sign}(y_0)\cdot a_y\left(\frac{|y_0|}{R_{y,\,\textrm{\small{max}}}}\right)^{b_y},\cr\cr
&&v_z({\cal X}_0)=\tanh\eta_{s0},
\label{granvelo}
\end{eqnarray}
where the function $\textrm{Sign}(x)\!=\!\pm1$ for positive/negative $x$ ensuring that the source is
transversely expansive. The magnitude of $v_x$~($v_y$) increases with $|x_0|$~($|y_0|$), with the
rate of increase determined by the positive parameters $a_x$ and $b_x$~($a_y$ and $b_y$).
In the calculations, we take $R_{x,\,\textrm{\small{max}}}/R_x\!=\!R_{y,\,\textrm{\small{max}}}/R_y\!=\!3$,
and consider the source elements initiated from the elliptical transverse region
$(x_0/R_{x,\,\textrm{\small{max}}})^2\!+\!(y_0/R_{y,\,\textrm{\small{max}}})^2\!<\!1$.
In this way, the requirement $|\textbf{\textit{v}}|<1$ will be naturally guaranteed with the parameters
$a_x$ and $b_x$~($a_y$ and $b_y$) appropriately chosen~\cite{Ru:2017nkc}.

With the temporal distribution of each source element being Gaussian,
the CMF space-time distribution of a source element~(SE) initiated from ${\cal X}_0$,
denoted $\rho^{\tiny{\textrm{SE}}}_{{\cal X}_0}$,
can be expediently written in the $X\!=\!(t,\rr)$ representation
as $\rho^{\tiny{\textrm{SE}}}_{{\cal X}_0}(X)=
\widetilde{\rho}^{\,\tiny{\textrm{SE}}}_{{\cal X}_0}(X\!-\!X_0)\!=\!\widetilde{\rho}^{\,\tiny{\textrm{SE}}}_{{\cal X}_0}(t-t_0, \,\rr-\rr_0)$, with
\begin{equation}
\widetilde{\rho}^{\,\tiny{\textrm{SE}}}_{{\cal X}_0}(X)
=\sqrt{\frac{2}{\pi}}\,T_s^{-1}
\exp\left(-\frac{t^2}{2T^2_s}\right)\,\delta^{(3)}(~\rr-\textbf{\textit{v}} t~),~~~~(t>0),
\label{rho0CMF}
\end{equation}
where the wide-tilde $\widetilde{\rho}^{\,\tiny{\textrm{SE}}}_{{\cal X}_0}(X)$ is the equivalent
source-element distribution with the initial coordinate shifted from $(t_0,\rr_0)$ to $(0,\textbf{{0}})$,
and $T_s$ is the CMF duration time of the source element~\cite{Ru:2017nkc}.
By considering that all the source elements have the same longitudinal proper lifetime $\tau_s$,
we have $T_s\!=\!\tau_s\!\cosh\eta_{s0}$.
For legibility, we supplement that $X_0=(t_0, x_0, y_0, z_0)=(\tau_0\!\cosh\eta_{s0}, x_0,\,y_0,\tau_0\!\sinh\eta_{s0})$.
Overall, $\widetilde{\rho}^{\,\tiny{\textrm{SE}}}_{{\cal X}_0}(X)$ depends on both $T_s\!=\!T_s({\cal X}_0)$ and the
source element moving velocity $\textbf{\textit{v}}=\textbf{\textit{v}}({\cal X}_0)$ in the form of Eq.~(\ref{granvelo}).

Assuming that all the source elements are evolved from the initial source $\rho_{\rm init}({\cal X}_0)$,
one can finally write the space-time distribution of the expanding coherent source in the CMF by integrating all the
sub-distributions $\rho^{\tiny{\textrm{SE}}}_{{\cal X}_0}(X)$ together as
\begin{equation}
\rho(X)=\int\! d^4\!{\cal X}_0\, \rho_{\rm init}({\cal X}_0)\,\rho^{\tiny{\textrm{SE}}}_{{\cal X}_0}(X),
\label{rhorho0}
\end{equation}
where $\int\! d^4\!{\cal X}_0\!\equiv\!\int\! d\tau_0dx_0dy_0d\eta_{s0}$ with normalization condition
$\int\! d^4\!{\cal X}_0\, \rho_{\rm init}({\cal X}_0)\!=\!1$.
At this point, we have completed the parametrization of the coherent source with both transverse and
longitudinal expansions.

The single-pion momentum distribution for the coherent source $\rho(X)$ can already be calculated
utilizing Eqs.~(\ref{FTclas}), (\ref{cohpoint}), and (\ref{P_c}). All the same, it is very useful to
introduce the pion-emission amplitude of the sub-source $\rho^{\tiny{\textrm{SE}}}_{{\cal X}_0}(X)$~\cite{Ru:2017nkc},
expressed as
\begin{eqnarray}
{\cal A}^{^{\tiny{\textrm{SE}}}}_{^{C}}({\cal X}_0, \pp)
\equiv\,A_{_C}(\pp)\,\,G_0\!\left\{\left[p\!\cdot u({\cal X}_0)\right](\gamma_u^{-1}T_s)\right\}
\,\,\,\,\,\,\,\,\,\,\,\,\,\,\,\,\,\,\,\,\,\,\,\,\,\,\,\,\,\,\,\,\, \cr\cr
=\!A_{_C}(\pp)\,\!\int\! d^4X~e^{ip\cdot X}\,\,\widetilde{\rho}^{\,\tiny{\textrm{SE}}}_{{\cal X}_0}(X)\,\,\,\,\,\,\,\,\,\,\,\,
\,\,\,\,\,\,\,\,\,\,\,\,\,\,\,\,\,\,\,\,\,\,\,\,\,\,\,\,\,\,\,\,\,\,\,\,\,\,\,\,\,\,\,\,\,\,\,\,\,\,\,\,\,\,\,\,\,\,\,\,
\cr\cr
=\!\left[2E_p(2\pi)^3\right]^{-\frac{1}{2}}\!\!\!
\int_{0}^\infty \!\!\!\!dt\!\sqrt{\frac{2}{\pi}} \exp\left[it\left(\left[p\!\cdot\! u({\cal X}_0)\right](\gamma_u^{-1}\!T_s)\right)-\!\frac{t^2}{2}\right],\,\,
\label{subamplc}
\end{eqnarray}
where $A_{_C}(\pp)$ is in the form of Eq.~(\ref{cohpoint}), the function $G_0$ is the Fourier transform of
the sub-distribution $\widetilde{\rho}^{\,\tiny{\textrm{SE}}}_{{\cal X}_0}(X)$,
and $u({\cal X}_0)\!=\!\gamma_u\left(1,\textbf{\textit{v}}({\cal X}_0)\right)$ is the source element 4-velocity,
with the Lorentz factor $\gamma_u=(1-\textbf{\textit{v}}^2)^{-1/2}$.
\begin{figure*}[hbt]
\includegraphics[scale=0.26]{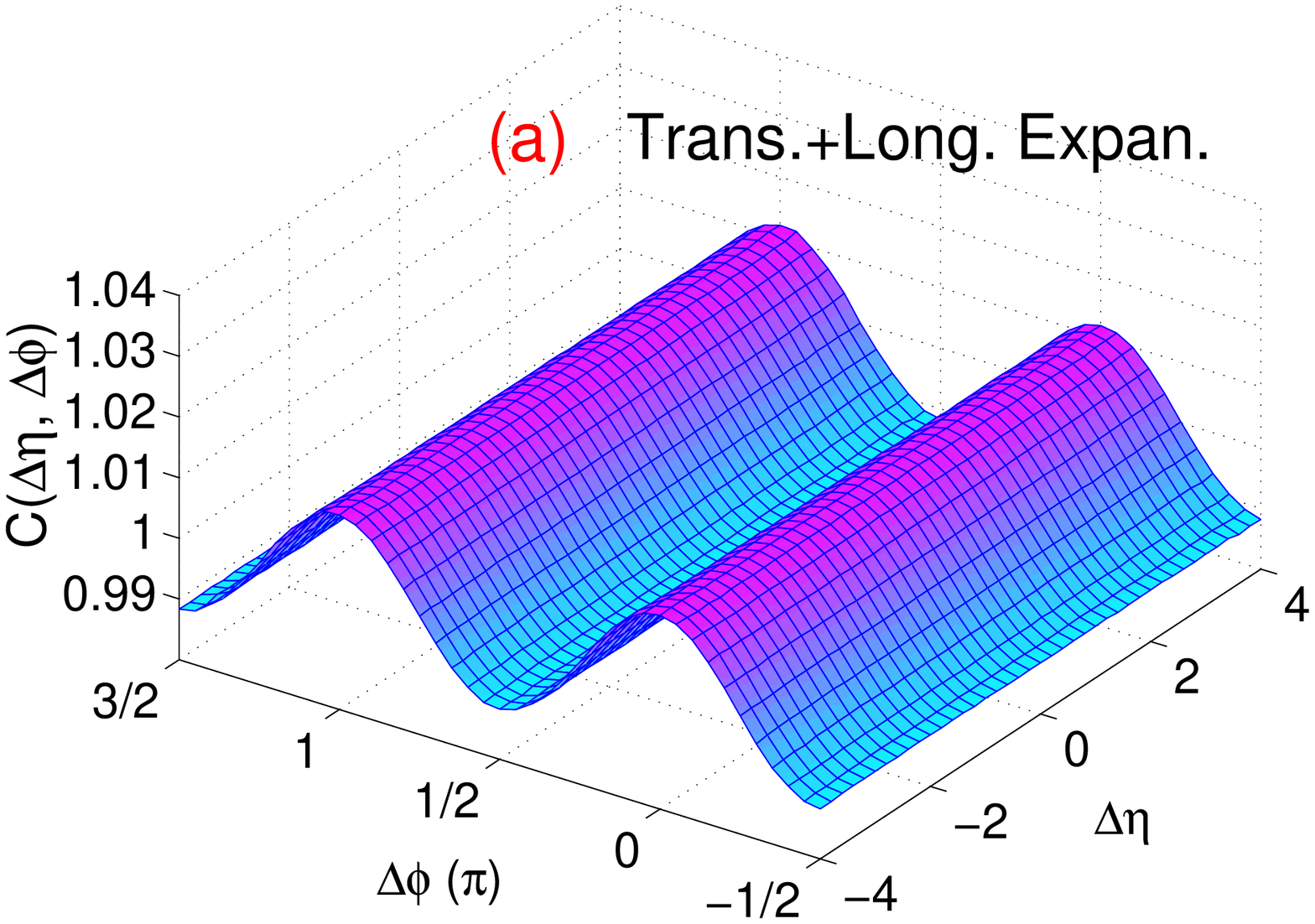}
\includegraphics[scale=0.26]{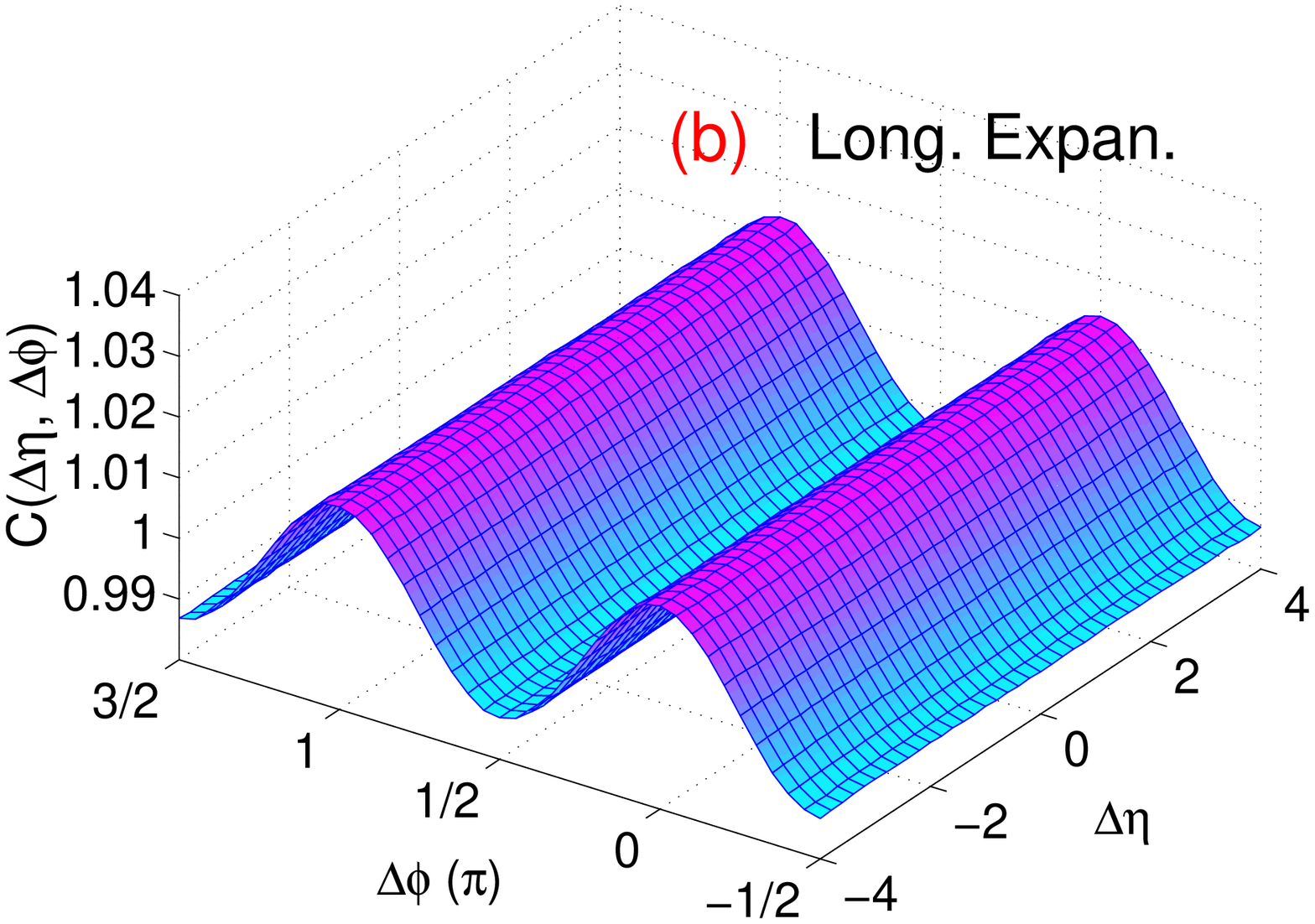}
\includegraphics[scale=0.26]{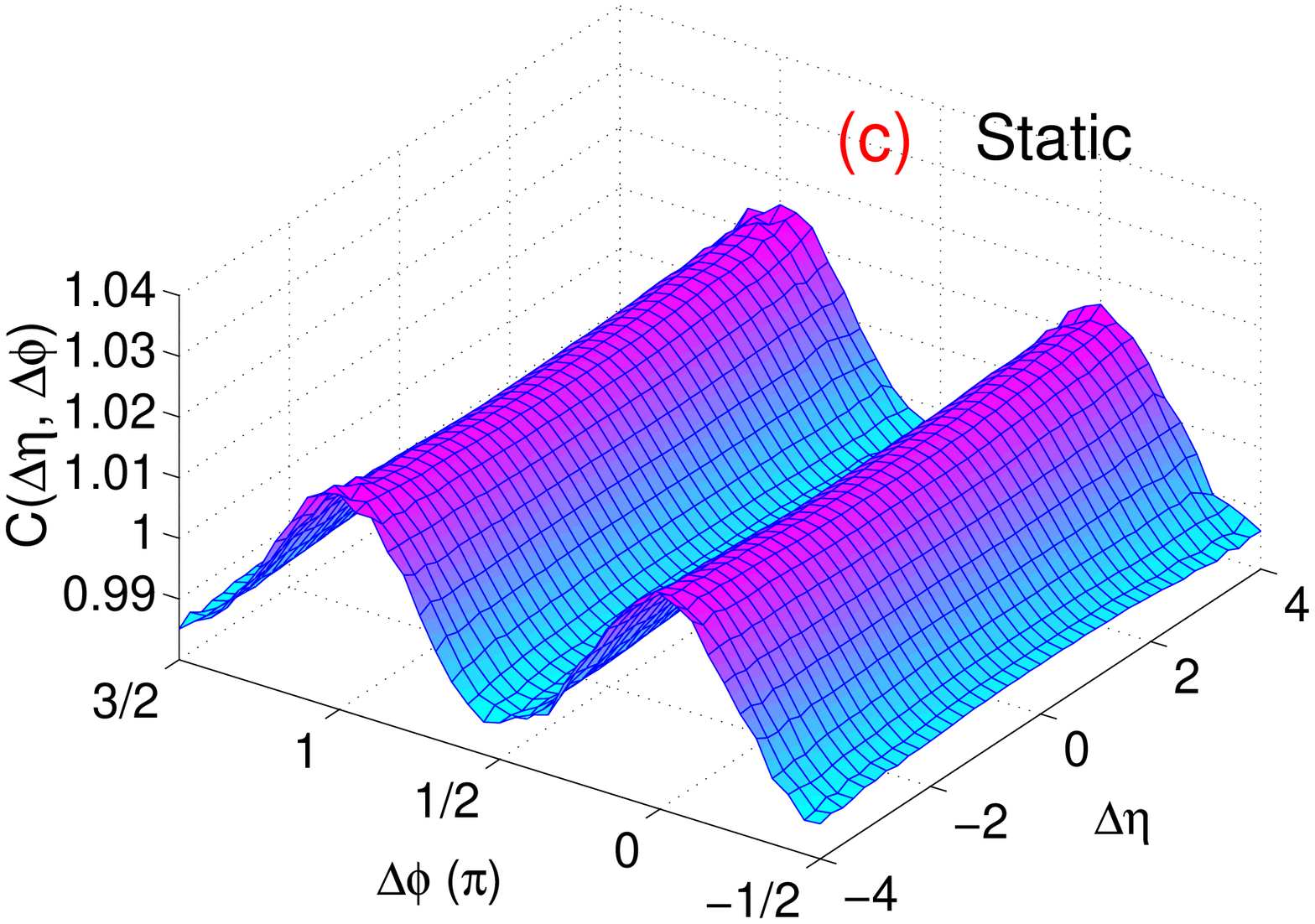}
\caption{(Color online) Two-particle correlation function, $C(\Delta\eta,\Delta\phi)$, for coherent pion-emitting sources,
with (a)~both transverse and longitudinal expansions, (b)~longitudinal expansion only, and (c)~no expansion~(static source).
In calculations, initial transverse size and shape of coherent source are taken to be
$R_T\!\equiv\!\sqrt{R_xR_y}\!=\!0.12$~fm and $S_T\!\equiv\!R_y/R_x\!=\!1.45$, respectively.
Range of source space-time rapidity is taken to be $\Delta\eta_{s0}=4$.
Initial time and lifetime of coherent source are taken to be $\tau_i\!=\!0$~fm/c and
$\tau_s\!=\!0.5$~fm/c, respectively.
Velocity parameters are taken to be $a_x\!=\!0.6$, $a_y\!=\!0.56$ and $b_{x,y}\!=\!0.5$~\cite{Ru:2017nkc} for
an anisotropic transverse expansion of coherent source~[panel (a)].
Kinematic region of pions is $|\eta|<2.5$ and $0.5<p_T<5.0$~GeV~\cite{Aad:2015gqa,Aaboud:2016yar}.}
\label{cohc2}
\end{figure*}

In this way, the Lorentz invariant pion momentum distribution for the coherent source $\rho(X)$ can be readily written
as the result of the coherent superposition of all the sub-source pion-emission amplitudes~\cite{Ru:2017nkc} as
%\begin{widetext}
\begin{eqnarray}
E_pP_{\!_C}(\pp)=E_p\left|\int\! d^4\!{\cal X}_0\,\rho_{\rm init}({\cal X}_0)\,e^{ip\cdot X_0}
{\cal A}^{^{\tiny{\textrm{SE}}}}_{^{C}}({\cal X}_0, \pp)\,\right|^2\,\,\,\,\,\,\,\,\,\,\,\,\,\,\,\,\,\,\cr\cr
=\!\left[2(2\pi)^3\right]^{-1}\!\left|\int\!\! d^4\!{\cal X}_0\,\rho_{\rm init}({\cal X}_0)\,
e^{ip\cdot X_0} G_0\!\left\{\left[p\!\cdot\!u({\cal X}_0)\right]\!(\gamma_u^{-1}T_s)\right\}\,\right|^2\!\cr\cr
=\!\left[2(2\pi)^3\right]^{-1}\bigg|\int\!\! d^4\!{\cal X}_0\,\rho_{\rm init}({\cal X}_0)\,
\exp\left[im_T\tau_0\cosh(y\!-\!\eta_{s0})\!-\!i\small{\pp_T\!\cdot\!{\hbox{\boldmath{$ x $}}}_{0T}}\right]\cr
\times \,{G_0\!\left\{\left[p\!\cdot\!u({\cal X}_0)\right]\!(\gamma_u^{-1}\!\!\cosh\eta_{s0}\tau_s)\right\}}\bigg|^2,\,\,\,\,\,\,\,\,\,\,\,\,\,\,\,\,\,\,\,\,
\,\,\,\,\,\,\,\,\,\,\,\,\,\,\,\,\,\,\,\,\,\,\,\,\,\,\,\,\,\,\,\,\,
\label{P_c2}
\end{eqnarray}
where $m_T\!=\!(p_T^2+m_{\pi}^2)^{1/2}$ is the pion transverse mass, $y$ the pion rapidity,
and ${\hbox{\boldmath{$ x $}}}_{0T}\!=\!(x_0, y_0)$ the transverse vector.
With the role of the source expansion involved in function $G_0$, Eq.~(\ref{P_c2}) reveals how the
momentum distribution is related to the source initial geometry featured in $\rho_{\rm init}$.
Equation~(\ref{P_c2}) will be used in the rest of this paper to study the long-range azimuthal correlations.
%\end{widetext}

\subsection{Long-range two-particle azimuthal correlations}
\label{c2}
Next, we study the two-particle angular correlation function defined as
\begin{eqnarray}
C(\Delta\eta,\Delta\phi)=\frac{S(\Delta\eta,\Delta\phi)}{B\,(\Delta\eta,\Delta\phi)},
\end{eqnarray}
which is commonly used in the experimental measurement~\cite{Aad:2015gqa,Aaboud:2016yar}.
In our theoretical calculations, to obtain the two-particle distribution $S(\Delta\eta,\Delta\phi)$,
we randomly generate $10^6$ pions in terms of the single-particle momentum distribution expressed as
Eq.~(\ref{P_c2}) and sort the pion pairs into kinematic bins defined with two-particle relative azimuthal
angle $\Delta\phi\!\equiv\!\phi_1\!-\!\phi_2$ and relative pseudo-rapidity $\Delta\eta\!\equiv\!\eta_1\!-\!\eta_2$.
Then, we similarly evaluate the ``background" distribution, $B(\Delta\eta,\Delta\phi)$, by additionally
imposing an azimuthal isotropy condition when randomly generating pions.
In this way, azimuthal correlations are eliminated in the obtained $B(\Delta\eta,\Delta\phi)$, similar
to the experimental measurement with ``mixed events"~\cite{Aad:2015gqa,Aaboud:2016yar}.
Therefore, the case $C(\Delta\eta,\Delta\phi)\!=\!1$ corresponds to a vanishing two-particle azimuthal correlation.

In panel (a) of Fig.~\ref{cohc2}, we show the two-particle angular correlation function for a
coherent pion-emitting source with both transverse and longitudinal expansions.
One can observe a remarkable double-ridge structure~\cite{Khachatryan:2016txc,Schenke:2016lrs}
for the coherent pion emission. To trace the origin of this phenomenon and to study the effects of the source expansion,
we show the results for two other typical cases, i.e., the source with longitudinal expansion only~(non-expansive in transverse)
and the source being static,
in panels (b) and (c) of Fig.~\ref{cohc2}, respectively. The source parameters for the three cases in
Fig.~\ref{cohc2} are same, except for the expansion velocity.
It is interesting to observe that the two-particle correlation functions for the three cases are very similar.
This implies that, for a coherent emission, the near-side ridge structure can arise without source collective expansion,
and the effect of the source expansion velocity on $C(\Delta\eta,\Delta\phi)$ is small~(also checked
by varying the transverse velocity and geometry).
It is noteworthy that the expressions of the pion momentum distributions for the latter two cases can be
simplified as
\begin{widetext}
\begin{eqnarray}
\label{Lexp}
\frac{dN_{^C}^3}{p_Tdp_Tdyd\phi}
=E_p\left|\int\! d{\hbox{\boldmath{$ x $}}}_{0T}\,\frac{(R_xR_y)^{-1}}
{2\pi}e^{i\small{\pp_T\cdot{\hbox{\boldmath{$ x $}}}_{0T}}}\exp{\left(-\frac{x_0^2}{2R_x^2}
-\frac{y_0^2}{2R_y^2}\right)}\int\!\! d\tau_{0}\!\!\int_{-\Delta\eta_{s0}}^{+\Delta\eta_{s0}}\!\!\!\! d\eta_{s0}\,\frac{\delta(\tau_0-\tau_i)}{2\Delta\eta_{s0}}\,
e^{im_T\tau_0\cosh(y-\eta_{s0})} \,{\cal A}^{^{\tiny{\textrm{SE}}}}_{^{C}}({\cal X}_0, \pp)\,\right|^2 \cr\cr\cr
\!\!=\frac{(2\pi)^{-3}}{2{(2\Delta\eta_{s0})}^2}\exp\bigg\{\!-p_T^2\left[(R_x\cos\phi)^2\!+\!(R_y\sin\phi)^2\right]\bigg\}
\,\left|\int_{-\Delta\eta_{s0}}^{+\Delta\eta_{s0}}\!\!\!\! d\eta_{s0}\,e^{im_T\tau_i\cosh(y-\eta_{s0})}
\,G_0\big(m_T\cosh(y-\eta_{s0})\tau_s\big)\right|^2,~~\cr
\textrm{(longitudinally expansive)}. \\\cr
\frac{dN_{^C}^3}{p_Tdp_Tdyd\phi}
=\frac{(2\pi)^{-3}}{2}\exp\bigg\{\!-p_T^2\left[(R_x\cos\phi)^2\!+\!(R_y\sin\phi)^2\right]\bigg\}
\,\,\bigg|G_0\big(m_T\cosh y\,\tau_s\big)\bigg|^2,~~~~~~~~~~~~~~~\textrm{(static source, }\tau_i=0~\textrm{fm/c)}.
\label{cohstatic}
\end{eqnarray}
\end{widetext}
Clearly, Eqs.~(\ref{Lexp}) and (\ref{cohstatic}) share the same elliptical-type azimuthal angle component
$e^{\!-p_T^2\left[(R_x\cos\phi)^2\!+\!(R_y\sin\phi)^2\right]}$, which is independent
of the rapidity. This factorization property is responsible for the strong long-range correlations
arising with the asymmetric transverse profile, i.e., $R_x\!\neq\!R_y$.
As discussed in Secs.~\ref{CS} and \ref{expansion}, this long-range azimuthal anisotropy is connected to the
source geometry through the interference in coherent emission, and can be viewed as an interference
pattern in momentum space.

\begin{figure}[t]
\includegraphics[scale=0.62]{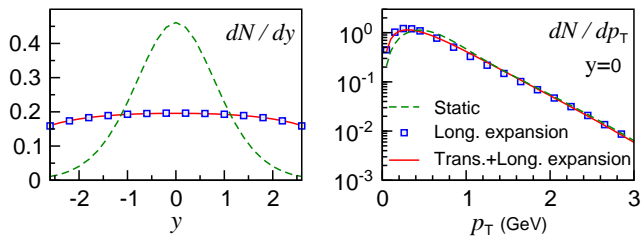}
\caption{(Color online) Normalized pion rapidity distribution~(left-hand panel) and transverse momentum distribution at central
rapidity~(right-hand panel) for coherent sources with both transverse and longitudinal expansions, with longitudinal
expansion only, and without expansion~(static source), corresponding to the results in Fig.~\ref{cohc2}. }
\label{cohexpan}
\end{figure}

To further study the effect of the source expansion on the coherent emission, we plot in Fig.~\ref{cohexpan}
the pion rapidity~(left-hand panel) and transverse momentum~(right-hand panel) distributions for the three typical cases.
One can see that the Bjorken longitudinal expansion of the coherent source has a significant impact on the rapidity
distribution, and a visible effect on the $p_T$ distribution. However, both the rapidity and the transverse momentum distributions
are insensitive to the transverse expansion of the coherent source.
This distinction is related to the asymmetry between the transverse and longitudinal source structures.
Due to the negligible effect of the transverse expansion, Eq.~(\ref{Lexp}) virtually serves as an approximation for the general case
involving both transverse and longitudinal source expansions, expressed as Eq.~(\ref{P_c2}).

The results presented in this section are consistent with the finding in our previous work~\cite{Ru:2017nkc},
where the effects of the transverse expansion of coherent source on the $p_T$ spectrum and $v_2$ are
shown to be largely reduced owing to the interference in coherent emission.
In contrast, the transverse expansion of chaotic source can generate considerable flow effects because of the
independent particle emissions at different space-time positions~\cite{Ru:2017nkc}.

To summarize, the long-range azimuthal correlations for the expanding coherent source
arise from the interference in coherent emission and are insensitive to the source collective expansion.
The correlation function $C(\Delta\eta,\Delta\phi)$ is to a large extent
related to the initial transverse profile of the source~(i.e., $R_x$ and $R_y$).

\section{Partially coherent pion emission in proton-proton collisions}
\label{part_coh}
\subsection{Chaotic component in partially coherent emission}
\label{chaotic}
In this subsection, we begin to focus on the long-range azimuthal correlations and the related phenomena
in pp collisions. In relativistic pp collisions, e.g., at the LHC, the created pion-emitting source
is possibly partially coherent~\cite{Aamodt:2011kd,Khachatryan:2011hi,Aad:2015sja,Aaij:2017oqu,Sirunyan:2017ies},
and there should be a chaotic component in the pion emission.
In general, the space-time structure and dynamical evolution of the chaotic source can be different from those of the coherent one.
To characterize the chaotic pion emission, we utilize the widely used blast-wave~(BW) spectrum~\cite{Schnedermann:1993ws,Abelev:2013vea}
as follows:
\begin{eqnarray}
\frac{dN_{\chi}^3}{p_Tdp_Tdyd\phi}
\propto\!\int_0^R\!\!r\,dr\,m_T I_0\left(\frac{p_T\sinh\rho}{T}\right)K_1\left(\frac{m_T\cosh\rho}{T}\right),~~
\label{BW}
\end{eqnarray}
where $I_0$ and $K_1$ are the modified Bessel functions, $T$ is the freeze-out temperature,
$R$ is the source radius, and $\rho$ is the transverse velocity profile given by
\begin{eqnarray}
\rho=\tanh^{-1}\left[\,\left(r/R\right)^n\beta_s \,\right],
\end{eqnarray}
with $\beta_s$ being the transverse expansion velocity at the surface.

\begin{figure}[t]
\includegraphics[scale=0.56]{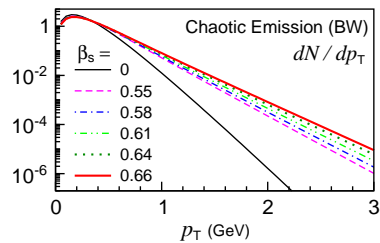}
\caption{(Color online) Normalized pion transverse momentum distribution for chaotic emission in blast-wave model.
In calculations, we take $T\!=\!100$~MeV and $n\!=\!2$.
Parameter $\beta_s$ for transverse expansion velocity of the chaotic source is taken to be 0, 0.55, 0.58,
0.61, 0.64, and 0.66.}
\label{BWpt}
\end{figure}

In Fig.~\ref{BWpt} we show the pion transverse momentum distributions for the chaotic pion-emitting source
with various transverse expansion velocities.
Obviously, the $p_T$ distribution becomes ``harder"~(wider) with increasing transverse expansion
velocity, which is expected and is usually referred to as the radial flow effect.

With the given momentum distributions for both coherent and chaotic emissions~[Eqs.~(\ref{P_c2}) and (\ref{BW})]
and their proportions in pion emission, the total distribution for a partially coherent source can be evaluated
by using Eq.~(\ref{parcoh}).
It should be mentioned that, to clarify the effect of coherent emission on the long-range
azimuthal correlations, in this work, we have not taken into account the anisotropic transverse expansion of the chaotic
source, which is able to generate anisotropic flow.
Concretely, in Eq.~(\ref{BW}) the single-particle momentum distribution for chaotic emission
is homogeneous for both rapidity and azimuthal angle components. Thus, in this
partially coherent pion emission model, the long-range azimuthal correlation will be entirely
from the coherent emission part.

\subsection{Extracting coherent fraction from two-pion HBT measurement}
\label{lambdaHBT}
\begin{figure}[t]
\includegraphics[scale=0.53]{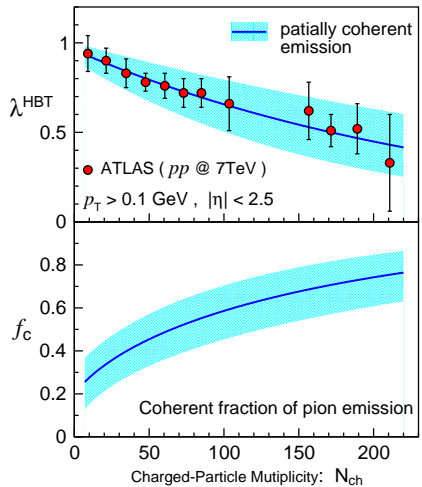}
\caption{(Color online) Strength parameter $\lambda^{\tiny{\textrm{HBT}}}$ of two-pion HBT correlation in pp
collisions at $\sqrt{s}\!=\!7$~TeV as a function of charged-particle multiplicity~(top panel)
and extracted coherent fraction of pion emission~(bottom panel).
In top panel, red disc represents ATLAS data~\cite{Aad:2015sja}, blue curve is fitted value with
parametrization form $\lambda^{\tiny{\textrm{HBT}}}(N_{\tiny{\textrm{ch}}})=\gamma e^{-N_{\tiny{\textrm{ch}}}\delta}$,
and shaded area corresponds to a likelihood band.}
\label{lambda}
\end{figure}

To complete the partially coherent source model, it is necessary to appropriately estimate the fraction of
coherent/chaotic emission in the total pion momentum distribution. As is discussed in Sec.~\ref{CS}, the pion
HBT correlations can provide an excellent probe of the degree of coherence in pion emission.
For example, the strength parameter $\lambda^{\tiny{\textrm{HBT}}}$ for the two-pion HBT correlations,
also called the chaoticity parameter, will decrease with increasing degree of coherence in pion emission.
Based on this, we extract the coherent fraction of pion emission, $f_c$, as a function of the charged-particle
multiplicity $N_{\tiny{\textrm{ch}}}$ from the measurement of $\lambda^{\tiny{\textrm{HBT}}}$
in pp collisions at $\sqrt{s}\!=\!7$~TeV performed by ATLAS~\cite{Aad:2015sja}, which is shown in Fig.~\ref{lambda}.

We note that the experimentalists have made numerous efforts to effectively exclude the final-state effects~\cite{Aad:2015sja},
e.g., the long-range Coulomb force, which may affect the measurement of $\lambda^{\tiny{\textrm{HBT}}}$. Therefore, we
assume that the measured $\lambda^{\tiny{\textrm{HBT}}}$ suppression~($\lambda^{\tiny{\textrm{HBT}}}<1$) is
to a large extent related to the presence of coherence in pion emission. In the top panel of Fig.~\ref{lambda},
the solid curve corresponds to the fitted values with the parametrization
form $\lambda^{\tiny{\textrm{HBT}}}(N_{\tiny{\textrm{ch}}})=\gamma e^{-N_{\tiny{\textrm{ch}}}\delta}$~\cite{Aad:2015sja}.
To address other effects that may smear the signal of coherence~\cite{Nickerson:1997js,Zhang:2009jw,Plumberg:2016sig},
e.g., the long-lived resonance decay,
we consider additionally a likelihood band around the fitted values shown as the shaded area.

With the extracted $\lambda^{\tiny{\textrm{HBT}}}(N_{\tiny{\textrm{ch}}})$, one can estimate
the coherent fraction in pion emission using the following relation~\cite{Gyulassy:1979yi,Wong:2007hx}:
\begin{eqnarray}
f_c(N_{\tiny{\textrm{ch}}})\approx\left[1-\lambda^{\tiny{\textrm{HBT}}}(N_{\tiny{\textrm{ch}}})\right]^{1/2}.
\end{eqnarray}
The obtained $f_c(N_{\tiny{\textrm{ch}}})$, to be used as an input of the partially coherent pion emission model,
is shown in the bottom panel of Fig.~\ref{lambda}, with the shaded area corresponding to the band of the extracted $\lambda^{\tiny{\textrm{HBT}}}(N_{\tiny{\textrm{ch}}})$.
It is interesting to see that the degree of coherence increases with $N_{\tiny{\textrm{ch}}}$, which may
indicate that the coherence is more likely to arise from the state with a higher density of particle occupation~\cite{Blaizot:2011xf,Berges:2019oun}.
A similar multiplicity dependence for the HBT correlation strengths has been observed in the other experiments at the LHC~\cite{Aamodt:2011kd,Khachatryan:2011hi,Aad:2015sja,Aaij:2017oqu,Sirunyan:2017ies}.

In high energy hadronic collisions, the possible coherent emission may be related to the occurrence of Bose-Einstein condensate~\cite{Blaizot:2011xf,Begun:2013nga,Begun:2015ifa},
the initial-stage Glasma field~\cite{Schenke:2016lrs}, or the string fragmentation as in the Schwinger model of
2-dimensional QED~\cite{Wong:2009eu}, etc.
The real mechanism is not yet clear.
A comprehensive study of the pion HBT correlations~\cite{Wong:2007hx,Bary:2018sue,Bary:2019kih}
and other relevant observables such as the single-particle momentum distribution~\cite{Begun:2013nga,Begun:2015ifa,Ru:2017nkc}
and $m$-particle azimuthal correlations~(e.g., $v_2\{m\}$) may provide more information on that, e.g.,
the space-time structure of the coherent source and the degree of coherence in pion emission.
In particular, high-order~(i.e., $m$-particle with $m>2$) HBT correlations and azimuthal
correlations should have different sensitivities to the coherent fraction
relative to the two-particle correlations. We will study these issues in a future work.

\begin{figure}[t]
\includegraphics[scale=0.53]{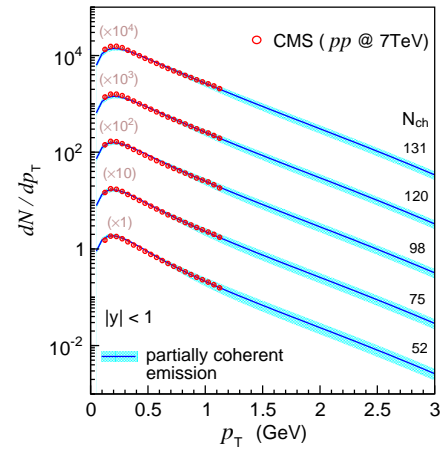}
\caption{(Color online) Normalized transverse momentum distribution of pion for five multiplicity classes in pp
collisions at $\sqrt{s}\!=\!7$~TeV.
Blue curve represents result of partially coherent emission, with shaded area corresponding
to band of extracted $f_c$ in Fig.~\ref{lambda}.
CMS data~\cite{Chatrchyan:2012qb} are shown as a red circle.
Source parameters are taken the same as in Figs.~\ref{cohc2} and \ref{BWpt}.}
\label{ptspec}
\end{figure}

\subsection{Results of partially coherent source}
\label{partial}
Next, we present the results of the partially coherent pion-emitting source for pp collisions
at $\sqrt{s}\!=\!7$~TeV.

Figure~\ref{ptspec} shows the results for pion transverse momentum distribution, in which the shaded areas
correspond to the bands of the extracted $f_c$ as shown in Fig.~\ref{lambda}.
One can see that the results of the partially coherent pion emission can well
describe the CMS data~\cite{Chatrchyan:2012qb} in the five multiplicity classes.

\begin{figure}[b]
\includegraphics[scale=0.53]{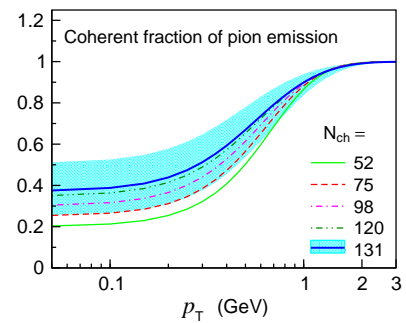}
\caption{(Color online) Coherent fraction of pion emission as a function of transverse momentum,
corresponding to results in Fig.~\ref{ptspec}.
Shaded area for $N_{\tiny{\textrm{ch}}}=131$ is from band of extracted $f_c$ in Fig.~\ref{lambda}.}
\label{coh_frac}
\end{figure}

In the calculations, for the coherent component, we consider a source with both longitudinal and transverse
expansions.
It is observed from the CMS data that the $p_T$ spectrum becomes harder
with increasing multiplicity~\cite{Chatrchyan:2012qb}, which may indicate a stronger radial flow effect
in the chaotic emission for a higher multiplicity.
Thus, we use a transverse expansion velocity increasing with the multiplicity for the chaotic source~(
the used values of $\beta_s$ are shown in Fig.~\ref{BWpt}).
We do not take into account the multiplicity dependence for the transverse expansion of the coherent source
due to the negligible observable effect as is discussed in Sec.~\ref{c2}.
To further simplify the model setting, we mainly focus on the events in the intermediate-to-high-multiplicity
region~(e.g., $50\!\lesssim\!N_{\tiny{\textrm{ch}}}\!\lesssim\!130$) in this work.
In total, there are two multiplicity-dependent physical quantities taken into account. One is the coherent fraction
in pion emission increasing with $N_{\tiny{\textrm{ch}}}$, which gives a decreasing two-pion HBT correlation strength.
The other is the transverse expansion velocity of the chaotic source increasing with $N_{\tiny{\textrm{ch}}}$, which
yields an increasing radial flow effect.

\begin{figure}[t]
\includegraphics[scale=0.53]{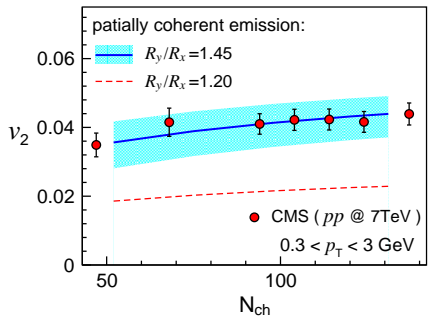}
\caption{(Color online) Elliptic anisotropy of pion as a function of charged-particle multiplicity
in pp collisions at $\sqrt{s}\!=\!7$~TeV.
Blue solid curve is result of partially coherent emission with $S_T\!=\!1.45$,
and shaded band is from extracted $f_c$ in Fig.~\ref{lambda}.
Red dashed curve is result with $S_T\!=\!1.20$.
Other source parameters are the same as in Fig.~\ref{ptspec}.
Red disc represents CMS $v_2^{\tiny\textrm{sub}}\{2\}$ data for charged particles~\cite{Khachatryan:2016txc}.}
\label{v2N}
\end{figure}

To better understand the pion transverse momentum distribution of the partially coherent emission, we plot in
Fig.~\ref{coh_frac} the coherent fraction of pion emission as a function of $p_T$.
We observe that the coherent fraction increases with $p_T$, which is
consistent with the experimental observation of a decreasing $\lambda^{\tiny{\textrm{HBT}}}$ versus
$p_T$~\cite{Khachatryan:2011hi,Aad:2015sja,Sirunyan:2017ies}.
It is interesting to note that a coherent fraction decreasing with $p_T$ was found in AA collisions
in our previous work~\cite{Ru:2017nkc}. The reason for this distinction is that, in AA collisions, a
stronger radial flow is expected to present in the chaotic emission; while, as the result of interferences,
the spectrum of coherent emission may be softer due to the larger transverse size of the coherent source~\cite{Ru:2017nkc}.

In Fig.~\ref{v2N}, we plot the pion elliptic anisotropy $v_2\!\equiv\!\langle(p_x^2\!-\!p_y^2)/p_T^2\rangle$
for the partially coherent emission as a function of the charged-particle multiplicity as the blue solid
curve. To illustrate the dependence on the initial transverse shape
of the coherent source for $v_2$, we also show the result with $S_T\!=\!1.20$ as the red dashed curve.
In general, $v_2$ will increase with both the transverse asymmetry of coherent source and
the coherent fraction~(or $N_{\tiny{\textrm{ch}}}$) in our partially coherent source model.
For comparison, we also show the CMS data for charged particles~\cite{Khachatryan:2016txc}.
We can see that, with the current model, the result for $S_T\!=\!1.45$ agrees well with the data.

\begin{figure}[h]
\includegraphics[scale=0.22]{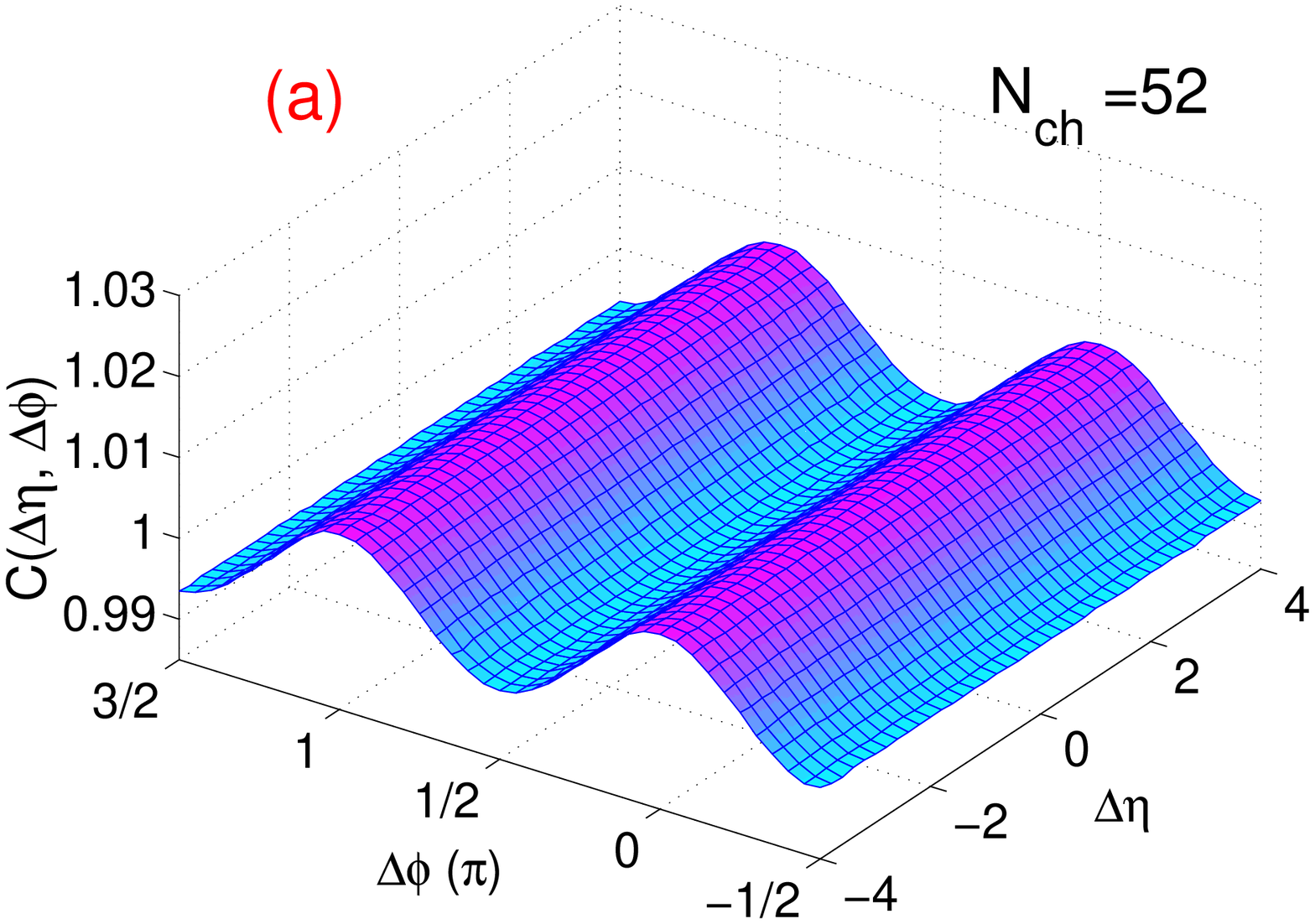}
\includegraphics[scale=0.22]{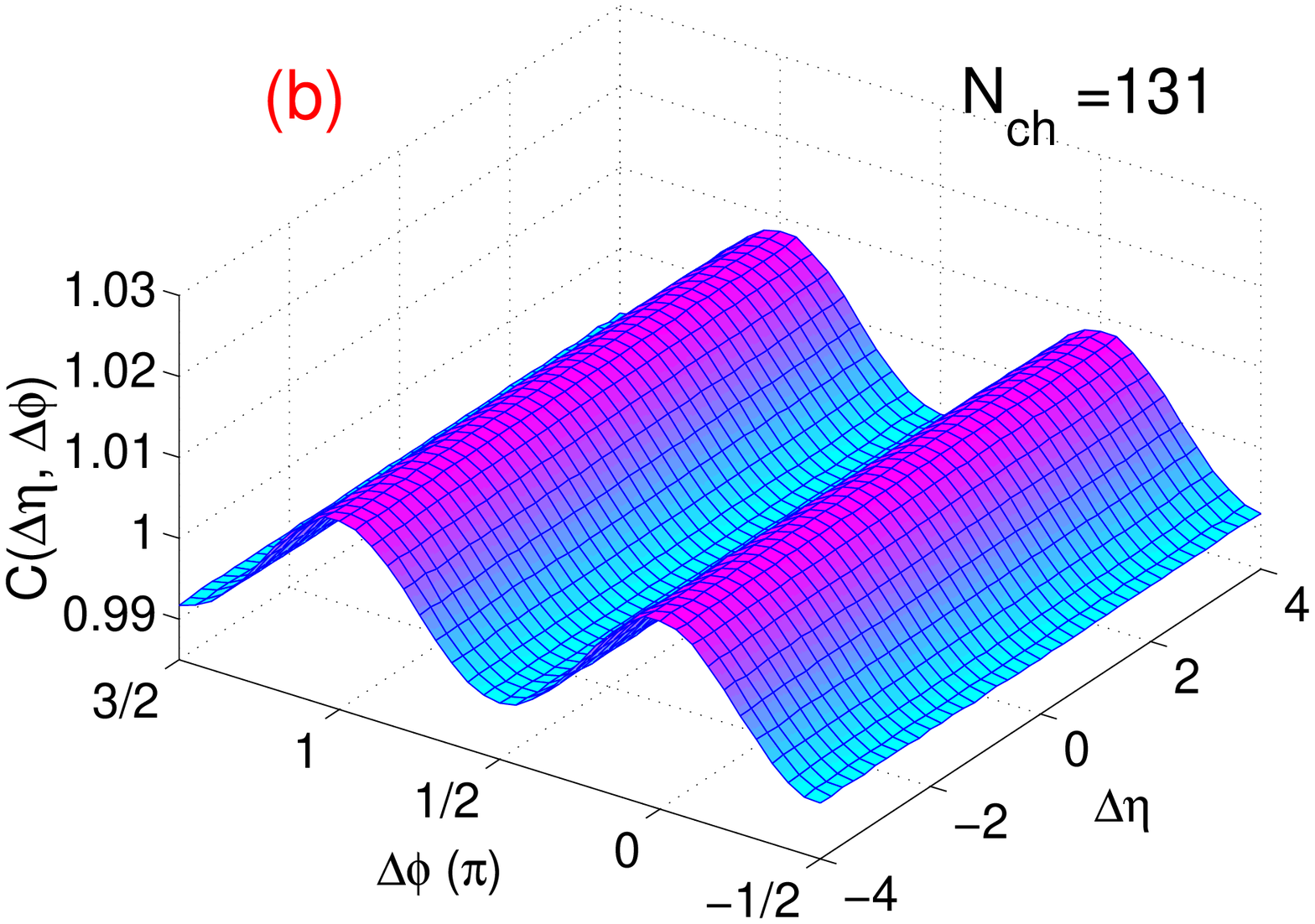}
\includegraphics[scale=0.52]{c2_eta4m.eps}
\caption{(Color online) Two-particle correlation function, $C(\Delta\eta,\Delta\phi)$,
for partially coherent pion-emitting sources for two multiplicity classes in pp collisions.
Panels (c) and (d) show values at $|\Delta\eta|=4$, with shaded band from extracted $f_c$ in Fig.~\ref{lambda}.
Source parameters are the same as in Fig.~\ref{ptspec}.
Kinematic region of pion is taken to be $|\eta|<2.5$
and $0.5<p_T<5.0$~GeV~\cite{Aad:2015gqa,Aaboud:2016yar}.}
\label{c2N}
\end{figure}

At the end of this section, we calculate the two-particle angular correlation function $C(\Delta\eta,\Delta\phi)$
of the partially coherent pion emission for pp collisions with $N_{\tiny{\textrm{ch}}}\!=$52 and 131, and show
the results in panels (a) and (b) of Fig.~\ref{c2N}, respectively.
For better comparison, the values of $C(\Delta\eta,\Delta\phi)$ at $|\Delta\eta|=4$ are plotted in panels
(c) and (d).
We observe from Fig.~\ref{c2N} that, with the presence of coherence in pion emission, the ridge structure
can manifest itself in the $C(\Delta\eta,\Delta\phi)$, and the correlation becomes stronger for a higher
multiplicity class, which corresponds to a larger coherent fraction in pion emission.
The results suggest that to deliberate the possible coherence in particle emission may be important for fully
understanding the collectivity in small systems.

~

\section{Summary and discussion}
\label{conclusion}
In this work, we investigate the influence of coherent pion emission on the long-range azimuthal
correlations in relativistic proton-proton collisions.

We study the pion momentum distribution for a coherent pion-emitting source with both transverse and longitudinal
expansions, and calculate the two-particle correlation function $C(\Delta\eta,\Delta\phi)$.
It is found that the function $C(\Delta\eta,\Delta\phi)$ has a remarkable ridge structure for the
coherent source whether it is expanding or static.
The onset of this long-range azimuthal correlation can be traced back to the asymmetric initial
transverse profile of the coherent source, owing to interference in coherent emission.
Because of the interference effect, the transverse expansion of the coherent
source has a slight influence on the pion momentum distribution. However, the
Bjorken longitudinal expansion significantly shapes the pion rapidity distribution.

To address the pion emission in pp collisions, we construct a partially coherent pion-emitting source
by incorporating the coherent source and a chaotic emission source described with the blast-wave model.
Using the experimental data of the two-pion HBT correlation strengths in pp collisions at $\sqrt{s}\!=\!7$~TeV,
we extract a coherent fraction of pion emission as an input of the partially coherent source model,
which increases with charged-particle multiplicity.
We find the results with the current model can well reproduce the experimental data of the
transverse momentum spectrum and the elliptic anisotropy in the intermediate-to-high multiplicity classes.
Furthermore, the ridge structure is still clear in the function $C(\Delta\eta,\Delta\phi)$
for the partially coherent emission. This long-range correlation becomes stronger with increasing
multiplicity due to the increasing degree of coherence.

It should be noted that, to clarify the effect of coherent emission on the $C(\Delta\eta,\Delta\phi)$,
we have taken no account of the possible emergence of the ridge correlation in the chaotic emission
component in this work, which may be related to the dynamics of the collective expansion of chaotic source,
and be more pronounced in higher-multiplicity events.
In the future, a more comprehensive study with various types of observable~\cite{Loizides:2016tew} should be helpful to refine the model.
Even so, the results in this paper indicate the importance of deliberating the possible coherent particle
emission, for fully understanding the collectivity in the small systems.

\begin{acknowledgments}
P.R. would like to thank Prof.~Ben-Wei Zhang for his kind invitation to visit CCNU.
This research was supported in part by the National Natural Science Foundation of China under
Grant No.~11675034, by the China Postdoctoral Science Foundation under Project No.~2019M652929,
by the MOE Key Laboratory of Quark and Lepton Physics~(CCNU) under Project No.~QLPL201802,
and by the Science and Technology Program of Guangzhou~(No.~2019050001).
\end{acknowledgments}

\end{document}